\documentclass[runningheads]{llncs}
\usepackage{amsfonts}
\usepackage{epsfig}
\usepackage{amsmath}
\usepackage{graphicx}
\usepackage{amsmath}
\usepackage{amscd}
\usepackage{stmaryrd}
\usepackage{algorithm}
\usepackage{arydshln}
\usepackage{algorithmic}
\usepackage{youngtab}
\usepackage{young}
\usepackage{geometry}
\usepackage{bbm}
\usepackage{leftidx}
\usepackage{qcircuit}

\usepackage[colorlinks,
            linkcolor=blue,
            anchorcolor=blue,
            citecolor=blue,
            ]{hyperref}

\def\bt{\begin{theorem}}
\def\et{\end{theorem}}
\def\bp{\begin{proposition}}
\def\ep{\end{proposition}}
\def\bc{\begin{corollary}}
\def\ec{\end{corollary}}
\def\bo{\begin{proof}}
\def\eo{\end{proof}}
\def\bx{\begin{example}}
\def\ex{\end{example}}
\def\br{\begin{remark}}
\def\er{\end{remark}}
\def\bl{\begin{lemma}}
\def\el{\end{lemma}}

\def\bn{\begin{definition}}
\def\en{\end{definition}}
\def\ba{\begin{array}}
\def\ea{\end{array}}
\def\be{\begin{equation}}
\def\ee{\end{equation}}
\def\bd{\begin{description}}
\def\ed{\end{description}}
\def\bu{\begin{enumerate}}
\def\eu{\end{enumerate}}
\def\bi{\begin{itemize}}
\def\ei{\end{itemize}}

\newbox\bigstrutbox
\setbox\bigstrutbox=\hbox{\vrule height12.5pt depth5pt width0pt}
\def\bigstrut{\relax\ifmmode\copy\bigstrutbox\else\unhcopy\bigstrutbox\fi}

\newbox\Bigstrutbox
\setbox\Bigstrutbox=\hbox{\vrule height17.5pt depth5pt width0pt}
\def\Bigstrut{\relax\ifmmode\copy\Bigstrutbox\else\unhcopy\Bigstrutbox\fi}

\def\ds{\displaystyle}

\def\x{{\bf x}}
\def\y{{\bf y}}

\def\0{{\bf 0}}
\def\1{{\bf 1}}
\def\2{{\bf 2}}
\def\3{{\bf 3}}
\def\4{{\bf 4}}
\def\5{{\bf 5}}
\def\6{{\bf 6}}
\def\7{{\bf 7}}
\def\8{{\bf 8}}
\def\9{{\bf 9}}

\begin{document}

\pagestyle{headings}

\mainmatter

\title{Quantum speedup to some types of polynomial equations}

\titlerunning{Quantum speedup to some types of polynomial equations}

\author{Changpeng Shao \\
cpshao@amss.ac.cn}

\authorrunning{Changpeng Shao}

\institute{Academy of Mathematics and Systems Science, Chinese Academy of Sciences, Beijing 100190, China}

\maketitle

\begin{abstract}
  In this paper, we consider three types of polynomial equations in quantum computer: linear divisibility equation, which belongs to a special type of binary-quadratic Diophantine equation;
  quadratic congruence equation with restriction in the solution and exponential congruence equation in finite field.
  Quantum algorithms based on Grover's algorithm and Shor's algorithm to these problems are given.
  As for the exponential congruence equation, which has been considered by Dam and Shparlinski \cite{dam} at 2008, a relatively simple quantum algorithm is given here.
  And some other results and generalizations are discovered.
\end{abstract}

\section{Introduction}

Quantum computer provide us more advantages to solve some problems efficiently than the classical Turing machine. Many efficient quantum algorithms towards some problems which are difficult to solve in the classical computer were discovered in the past decades (for example see the review \cite{childs}).

The polynomial equation solving problem is a typical hard but important problem with a lot of applications.
Different from numbers, polynomial is a little complicate to deal with in quantum computer now.
By Shor's work \cite{shor}, congruence equation can be solved efficiently in quantum computer.
The HHL algorithm \cite{harrow} seems to be another breakthrough related to this problem, although it can only give a state about the solution of linear system.
A work about diagonal equation with three terms in finite field was studied in \cite{dam}.
And recently, two works about polynomial interpolation in finite field were given in \cite{chen}, \cite{childs2}.

In this paper, we consider three types of polynomial equation solving problem, which seems cannot solved efficiently by the already existed quantum algorithms, especially Shor's algorithms.

The first equation is linear divisibility equation. It is a problem aims at finding a nontrivial factor of a positive integer $b$ with the form $ax+1$. This problem is a candidate of NPI problem \cite{adleman}.
It is a special type of binary-quadratic Diophantine equations, and was considered in \cite{adleman} to study P and NP problem.
It is proved to be $\gamma$-complete, but not known to be NP-complete.
A result in \cite{garey} shows that if
the linear divisibility equation can be solved in polynomial time, then NP $=$ co-NP.
Besides this, this problem has its own importance to study. Since it is a factoring problem, however, Shor's algorithm can only provide some partial helps.

The second equation studied in this paper is quadratic congruence equation with restriction in the solution. The general quadratic congruence equation $x^2\equiv a\mod b$ can be solved efficiently be Shor's  algorithms. However, it is a NP-complete problem \cite{manders} if we want to find a solution satisfies $0\leq x\leq c$. Some people believes that quantum computer cannot solve NP-complete problem efficiently (for example see \cite{aaronson}). This might be a problem to study the potential of quantum computer.

The third equation considered in the paper is about exponential congruence equation in finite field, i.e., diagonal equation studied in \cite{dam}.
The importance of this equation was discussed in paper \cite{dam}. However, here we care about this equation is because of its relations to polynomial equation in finite filed.

As for these three types of polynomial equations, Grover's algorithm and Shor's algorithm can only provide partial helps. We cannot rely on these two algorithms to solve these equations efficiently in quantum computer. As for the first two equations, we can achieve a cubic speedup in quantum computer based on the two algorithms.
As for the third equation, we present a relatively simple quantum algorithm to it compared with \cite{dam}. But the worst complexity is not improved here. And some generalizations of this equation and the relation to general polynomial equation in finite field of this equation were also discussed in this paper.

\section{Linear divisibility problem}
\setcounter{equation}{0}

The linear divisibility problem \cite{garey} states that for any two positive integers $a,b$, does there exists a positive integer $x$, such that $ax+1$ divides $b$?
This is equivalent to decide whether or not the following Diophantine equation
\be \label{eq1}
(ax+1)y=b
\ee
contains positive solutions for $x,y$?
In the following, we try to find quantum algorithm towards this problem. We first consider in the case when $a$ is prime.

By Shor's quantum factoring algorithm \cite{shor}, we can completely factor $b=p_1^{d_1}p_2^{d_2}\cdots p_n^{d_n}$ in time $O(\textmd{poly}(\log b))$. Then (\ref{eq1}) is equivalent to:
Does there exist $0\leq t_i\leq d_i~(i=1,2,\ldots,n)$, not all equal to zero, such that
\be \label{eq2}
p_1^{t_1}p_2^{t_2}\cdots p_n^{t_n}\equiv1\mod a.
\ee

Since $a$ is prime, so $\mathbb{Z}_a^*$ has a generator $g$, which can be computed in time $O(\textmd{poly}(\log a))$ \cite{dubrois}. By Shor's quantum discrete logarithm algorithm, we can computer $\lambda_i~(i=1,2,\ldots,n)$ in time $O(\textmd{poly}(\log a))$, such that
\[p_i\equiv g^{\lambda_i}\mod a.\]
Thus (\ref{eq1}) is equivalent to the following linear equation with restrictions:
\be \label{eq3}
\left\{
  \begin{array}{ll} \vspace{.2cm}
    \lambda_1t_1+\lambda_2t_2+\cdots+\lambda_nt_n\equiv0\mod (a-1), & \hbox{} \\ \vspace{.2cm}
    0\leq t_i\leq d_i,~~i=1,2,\ldots,n, & \hbox{} \\
    t_1+t_2+\cdots+t_n>0. & \hbox{}
  \end{array}
\right.
\ee
The total costs of the reduction from (\ref{eq1}) to (\ref{eq3}) is $O(\textmd{poly}(\log b))$.

In order to test whether (\ref{eq3}) contains a solution or not, a naive method is searching all possible cases of $t_i$. It contains $D=(d_1+1)(d_2+1)\cdots(d_n+1)$ cases. A classical result in number theory \cite{hua} states that for any $\epsilon$, we have $D=O(b^\epsilon)$, the omitted constant $A_\epsilon=(\frac{2}{\epsilon \ln(2)})^{2^{1/\epsilon}}$ in $O(b^\epsilon)$ depends on $\epsilon$. And when $\epsilon$ is small, $A_\epsilon$ can be very large. For example if $\epsilon=1/5$, then $A_\epsilon\approx 1.2\times 10^{37}$. Although $A_\epsilon$ is a constant, not good to be vary large in algorithm. Another classical result in number theory \cite{hua} states that for almost $b$, we have $D=O(\log b)$. So the average complexity of linear divisibility problem in quantum computer is $O(\textmd{poly}(\log b))$.

Direct searching by Grover's algorithm \cite{grover} only provides a quadratic speedup. However, based on the linear properties of (\ref{eq3}), we can actually achieve a cubic speedup, i.e., $O(b^{\epsilon/3}\textmd{poly}(\log b))$ in quantum computer, by the method introduced in \cite{brassard}. The basic idea is dividing $\{1,2,\ldots,n\}$ into two parts $\{1,2,\ldots,j\}\cup\{j+1,j+2,\ldots,n\}$, such that $(d_1+1)(d_2+1)\cdots(d_j+1)=O(D^{1/3})$. Then denote
\[S_1=\{\lambda_1t_1+\lambda_2t_2+\cdots+\lambda_jt_j\mod (a-1):0\leq t_i\leq d_i,1\leq i\leq j\},\]
and $S_2$ is the set by sorting $S_1$. The total complexity to get the set $S_2$ is $O(D^{1/3}\log D)$.
Then define a map $f$ as
\[\ba{rll} \vspace{.2cm}
f:\mathbb{Z}_{1+d_{j+1}}\times\cdots\times \mathbb{Z}_{1+d_n} &\rightarrow& \mathbb{Z}_2 \\
f(t_{j+1},\ldots,t_n) &=& \left\{
                            \begin{array}{ll} \vspace{.2cm}
                              0, & \hbox{if $\exists~v\in S_2$, such that $\lambda_{j+1}t_{j+1}+\cdots+\lambda_jt_j\equiv -v\mod (a-1)$;} \\
                              1, & \hbox{otherwise}
                            \end{array}
                          \right.
\ea\]
The value of the map $f$ can be computed by binary searching in the set $S_2$, which costs $O(\log D)$. And the way to find a solution of $f$ can be achieved by Grover's searching with complexity $O(D^{1/3})$. Therefore, the total complexity to solve (\ref{eq3}) is $O(b^{\epsilon/3}\textmd{poly}(\log b))$.

In the above searching procedure, equation (\ref{eq3}) actually does not provide too much help to solve (\ref{eq1}). Since the above searching procedure can be done directly in (\ref{eq2}) and this does not depends on whether $a$ is prime or not. And most importantly, when $a$ is not prime, (\ref{eq3}) will be changed into a linear system, which will be difficult to solve. At least, the above searching algorithm does not works anymore. Anyway, we have
\bp
The linear divisibility equation (\ref{eq1}) can be solved in quantum computer in time $O(b^{\epsilon/3}\emph{poly}(\log b))$ for some $\epsilon$ such that $A_\epsilon=(\frac{2}{\epsilon \ln(2)})^{2^{1/\epsilon}}$ is not too large.
\ep

\br
The linear divisibility problem provides us a new problem about factoring. Given $a,b,N$, does $N$ contains a factor in the form $ax+b$. The direct generalization of Shor's algorithm seems impossible to solve this problem efficiently. Also, as we can see in (\ref{eq3}), if $a$ is prime, then the equation $(ax+b)y=N$ is equivalent to
\be \label{eq30}
\left\{
  \begin{array}{ll} \vspace{.2cm}
    \lambda_1t_1+\lambda_2t_2+\cdots+\lambda_nt_n\equiv s\mod (a-1), & \hbox{} \\
    0\leq t_i\leq d_i,~~1\leq i\leq n. & \hbox{}
  \end{array}
\right.
\ee
for some $\lambda_1,\ldots,\lambda_n,s$. If $d_i=1$ for all $i$, then this is the subset sum problem, which brings new evidence to the difficulty of linear divisibility problem.
\er

\section{Quadratic congruence}
\setcounter{equation}{0}

With the finding of efficient factoring and discrete logarithm quantum algorithm, quadratic congruence equation can be solved efficiently in quantum computer. But quadratic congruence equation with restrictions on the solution cannot solved easily by Shor's work. In \cite{garey}, it introduces the following problem: For any three positive integers $a,b,c$, does there exists $0<x<c$, such that $x^2\equiv a\mod b$? This problem is proved to be NP-complete \cite{manders}. A basic idea to solve this problem in quantum computer is as follows:

We first factor $b=p_1^{d_1}p_2^{d_2}\cdots p_n^{d_n}$ by Shor's algorithm. Then all the solutions of $x^2\equiv a\mod b$ can be computed from
$x^2\equiv a\mod p_i^{d_i}$ for $1\leq i\leq n$. Let $\pm z_i$ be two solutions of $x^2\equiv a\mod p_i^{d_i}$, then by Chinese Reminder Theorem, the solution of $x^2\equiv a\mod b$ can be written as
\be\label{eq4}
x=\sum_{i=1}^n\mu_iz_i \alpha_i B_i \mod b,
\ee
where $\mu_i=\pm 1,B_i=b/p_i^{d_i}$ and $\alpha_i$ satisfies $\alpha_i B_i\equiv1 \mod p_i^{d_i}$. Therefore, the problem reduces to find suitable $\mu_i\in\{1,-1\}$ such that
\be
0<\sum_{i=1}^n\mu_iz_i \alpha_i B_i \mod b<c.
\ee
The method analyzed above to linear divisibly problem still works here and achieves a cubic speedup. Since $n<\log b$, so
\bp
There is a quantum algorithm that solves the problem whether or not there exists $0<x<c$, such that $x^2\equiv a\mod b$ in time $O(b^{1/3}\emph{poly}(\log b))$.
\ep

\br
This special type of quadratic congruence induces a more general problem as follows: Let $a_1,a_2,\ldots,a_n,b,c$ be some positive integers, does there exists $x_1,x_2,\ldots,x_n\in\{1,-1\}$, such that
\[0<a_1x_1+a_2x_2+\cdots+a_nx_n\mod b<c.\]
\er

\section{Exponential congruence equation}
\setcounter{equation}{0}

Let $\mathbb{F}_q$ be a finite field, $g$ be the generator of $\mathbb{F}_q^*$. Given $a,b,c,d\in\mathbb{F}_q^*$ and $a,c$ are factors of $q-1$, then does there exists $x,y\in \mathbb{F}_q$ satisfies the following equation in $\mathbb{F}_q$:
\be \label{eq5}
g^{ax+b}+g^{cy+d}=1.
\ee

A direct method is combining Grover's searching and Shor's discrete logarithm algorithm, just like the algorithm given in \cite{dam}. The method provided in \cite{dam} contains some hard estimations about the number of solutions of the equation (\ref{eq5}). In the following, we give a much simpler quantum algorithm to this problem compared with \cite{dam}. The worst complexity given in \cite{dam} does not improved here, but we get some different results from \cite{dam}. And some generalizations and observations about polynomial equation in finite field are discovered.

Assume that $a\leq c$. A result \cite{mullen} about the number $N$ of solutions of equation (\ref{eq5}) states that
\be
|N-q|\leq (a-1)(c-1)q^{1/2}<c^2q^{1/2}.
\ee
Hence, $N\geq q-c^2\sqrt{q}$.

Case 1, if $c\leq\ds\frac{1}{2}q^{1/4}$, then $\ds N\geq q-\frac{1}{4}q=\frac{3}{4}q$. Note that if $(x_0,y_0)$ is a solution of (\ref{eq5}), then there will exist $x_1,\ldots,x_{a-1}$, such that $(x_i,y_0)~(1\leq i\leq a-1)$ are solutions of (\ref{eq5}).
So in the searching space $\mathbb{F}_q$ of $y$, we have a probability at least $\ds\frac{3}{4}q/aq=3/4a$ to get the right $y$, such that by putting $y$ into (\ref{eq5}), we will get a $x$ such that $(x,y)$ is a solution of (\ref{eq5}). So direct searching will cost $\widetilde{O}(\sqrt{a})=\widetilde{O}(q^{1/8})$.

Case 2, if $q-c^2\sqrt{q}\leq 0$, i.e., $c\geq q^{1/4}$.
Since $a,c$ are factors of $q-1$, so we can set $q-1=as=ct$. Then we know that we only need to search $x$ and $y$ in space $\{0,1,\ldots,s-1\}$ and $\{0,1,\ldots,t-1\}$ respectively. Moreover, given any $y_0\in \mathbb{Z}_t$, if there is a $x_0\in\mathbb{Z}_s$ such that $(x_0,y_0)$ is a solution of (\ref{eq5}), then $x_0$ can be computed efficiently by Shor's algorithm. So we only need to search $y$ in space $\mathbb{Z}_t$ and then check whether or not the (\ref{eq5}) has a solution for $x$ by Shor's algorithm.
At this time, the complexity to find one solution of (\ref{eq5}) or decide whether (\ref{eq5}) has a solution is $\widetilde{O}(\sqrt{q/c})=\widetilde{O}(q^{3/8})$.
This result also holds for $c\geq \ds\frac{1}{2}q^{1/4}+1$.

\bp
Assume $a,b,c,d\in\mathbb{F}_q^*$ and $a\leq c$ are factors of $q-1$, then there is a quantum algorithm can decide whether or not the equation $g^{ax+b}+g^{cy+d}=1$ contains a solution in time $\widetilde{O}(\sqrt{q/c})$ if $c\geq \ds\frac{1}{2}q^{1/4}+1$, and in time $\widetilde{O}(\sqrt{a})$ if $c<\ds\frac{1}{2}q^{1/4}+1$. Moreover, if it has solutions, then with in the same complexity, the quantum algorithm will return one solution.
\ep

Besides the efficiency of this problem, we can do more generalizations about polynomial equation in finite field based on the above problem.
Note that (\ref{eq5}) is equivalent to diagonal equation $\lambda_1u^a+\lambda_2v^c=1$, here $\lambda_1=g^b,\lambda_2=g^d$ and $u=g^x,v=g^y$.
The general diagonal equation with $n$ terms in finite field $\mathbb{F}_q$ has the form
\be \label{eq7}
a_1x_1^{k_1}+a_2x_2^{k_2}+\cdots+a_nx_n^{k_n}=b,
\ee
where $a_1,\ldots,a_n,b\in \mathbb{F}_q^*$ and $k_1,\ldots,k_n\in\mathbb{Z}_{q-1}$. This equation can also written in the form (\ref{eq5}). If there is a $1\leq i\leq n$ such that $\gcd(k_i,q-1)=1$, then (\ref{eq7}) is trivial to solve by Shor's discrete logarithm algorithm. So we can assume $k_1,\ldots,k_n$ are factors of $q-1$ in (\ref{eq7}).

In the general case, the result in \cite{mullen} states that the number $N$ of solution of equation (\ref{eq7}) satisfies
\[|N-q^{n-1}|\leq(k_1-1)(k_2-1)\cdots(k_n-1)q^{(n-1)/2}<k_1k_2\cdots k_n q^{(n-1)/2}.\]
We can also assume $k_1\leq k_2\leq \cdots\leq k_n$, then $N\geq q^{n-1}-k_1k_2\cdots k_nq^{(n-1)/2}$. The following analysis is similar to the equation (\ref{eq5}).

Case 1, if $k_1k_2\cdots k_n<\ds\frac{1}{2}q^{(n-1)/2}$, then $N>\ds\frac{1}{2}q^{n-1}$. Since if $(x_{10},x_{20},\ldots,x_{n0})$ is a solution of equation (\ref{eq7}), then there exist $x_{11},\ldots,x_{1(k_1-1)}$, such that $(x_{1i},x_{20},\ldots,x_{n0})$ is a solution of equation (\ref{eq7}) for any $1\leq i\leq k_1-1$. The searching space of $(x_2,\ldots,x_n)$ is $\mathbb{F}_q^{n-1}$. So in $\mathbb{F}_q^{n-1}$, we have a probability larger than $\ds\frac{1}{2}q^{n-1}/k_1q^{n-1}=1/2k_1$ to get the right $x_2,\ldots,x_n$, such that by putting them into (\ref{eq7}), we will find a $x_1$ such that $(x_1,x_2,\ldots,x_n)$ is a solution of equation (\ref{eq7}). Since $k_1^n\leq k_1k_2\cdots k_n<\ds\frac{1}{2}q^{(n-1)/2}$, so the complexity is $\widetilde{O}(\sqrt{k_1})=\widetilde{O}(q^{(n-1)/4n})$.

Case 2, if $k_1k_2\cdots k_n\geq q^{(n-1)/2}$, then $N\geq 0$, which means equation (\ref{eq7}) may not contains a solution, so the above analysis does not works anymore. For any $i$, set $k_il_i=q-1$. Different from the above case, we only need to search each $x_i$ in the space $\mathbb{Z}_{l_i}$. And we only need to search $(x_2,\ldots,x_n)\in\mathbb{Z}_{l_2}\times \cdots\times \mathbb{Z}_{l_n}$, then check whether or not the equation (\ref{eq7}) has a solution for $x_1$. Therefore, at this time, the size of the searching space is
\[\frac{k_1q^{n-1}}{k_1k_2\ldots k_n}\leq \frac{k_1q^{n-1}}{q^{(n-1)/2}}=k_1q^{(n-1)/2}.\]
As we know, the quantum searching and discrete logarithm algorithm will provide a quadratic speedup to this case,
so the complexity is $\widetilde{O}(\sqrt{q^{n-1}/k_2\ldots k_n})=\widetilde{O}(\sqrt{k_1}q^{(n-1)/4})$. This also holds for $k_1k_2\cdots k_n\geq\ds\frac{1}{2}q^{(n-1)/2}$.

\bp
Assume $a_1,\ldots,a_n,b\in\mathbb{F}_q^*$ and $k_1\leq k_2\leq \cdots\leq k_n$ are factors of $q-1$. Then there is a quantum algorithm can decides whether or not the equation $a_1x_1^{k_1}+a_2x_2^{k_2}+\cdots+a_nx_n^{k_n}=b$ contains a solution in time $\widetilde{O}(\sqrt{q^{n-1}/k_2\ldots k_n})$ if $k_1k_2\cdots k_n\geq\ds\frac{1}{2}q^{(n-1)/2}$,
and in time $\widetilde{O}(\sqrt{k_1})$ if $k_1k_2\cdots k_n<\ds\frac{1}{2}q^{(n-1)/2}$. Moreover, if it has solution, then with in the same complexity, the quantum algorithm will return one solution.
\ep

\br
Equation (\ref{eq5}) can be written in the form $g^{cy}=g^{-d}-g^{ax+b-d}$. From the point of discrete logarithm, this problem is equivalent to find $x$ such that the discrete logarithm $l$ of $g^{-d}-g^{ax+b-d}$ in finite field $\mathbb{F}_q$ must divisible by $c$. So we can consider a general discrete logarithm problem, i.e., try to solve the equation
\be \label{eq6}
g^{cy}=f(x_1,\ldots,x_n)
\ee
in finite filed $\mathbb{F}_q$, here $c\mid (q-1)$ and $f$ is any polynomial belongs to $\mathbb{F}_q[x_1,\ldots,x_n]$. This problem aims at finding suitable $x_{10},\ldots,x_{n0}\in\mathbb{F}_q$, such that the discrete logarithm of $f(x_{10},\ldots,x_{n0})$ in finite field $\mathbb{F}_q^*$ is divisible by $c$.
Direct generalization of Shor's discrete logarithm seems impossible to this problem.
This is because the following problem cannot be solved efficiently in quantum computer due to the optimality of Grover's algorithm:
Let $F(t_1,\ldots,t_n)$ be any function over a set, such as $\mathbb{Z}_M^n$, and given a number $d\in\mathbb{Z}_M$, then does there exists $s_1,\ldots,s_n\in\mathbb{Z}_M$,
such that $d$ is a factor of $F(s_1,\ldots,s_n)$?
\er

In finite field $\mathbb{F}_q$, a single polynomial equation in $n$ variables can always be changed into the form
\be \label{eq8}
\lambda_1g^{a_{11}x_1+\cdots+a_{1n}x_n}+\lambda_2g^{a_{21}x_1+\cdots+a_{2n}x_n}+\cdots+\lambda_kg^{a_{k1}x_1+\cdots+a_{kn}x_n}=c.
\ee
We can introduce new variables to equation (\ref{eq8}), for any $1\leq j\leq k$, define
\be\ba{rll} \vspace{.2cm}
d_j    &:=& \gcd(a_{j1},\ldots,a_{jn}), \\
d_jy_j &:=& a_{j1}x_1+\cdots+a_{jn}x_n.
\ea\ee
Then equation (\ref{eq8}) is changed into
\be \label{eq9}
\lambda_1g^{d_1y_1}+\lambda_2g^{d_2y_2}+\cdots+\lambda_kg^{d_ky_k}=c.
\ee
Note that, if there is a $1\leq i\leq k$, such that $d_i=1$, then (\ref{eq9}) always has a solution in $\mathbb{F}_q$.

Let $y_1,y_2,\ldots,y_k$ be a solution of (\ref{eq9}), then in order to recover one solution of (\ref{eq8}), we need to solve the following linear Diophantine system
\be \label{eq10}
\left(
  \begin{array}{cccc}\vspace{.2cm}
    a_{11} &~~ a_{12} &~~ \ldots &~~ a_{jn} \\\vspace{.2cm}
    a_{21} &~~ a_{22} &~~ \ldots &~~ a_{2n} \\\vspace{.2cm}
    \vdots &~~ \vdots &~~ \ddots &~~ \vdots \\
    a_{k1} &~~ a_{k2} &~~ \ldots &~~ a_{kn} \\
  \end{array}
\right)\left(
         \begin{array}{c}\vspace{.2cm}
           x_1 \\\vspace{.2cm}
           x_2 \\\vspace{.2cm}
           \vdots \\
           x_n \\
         \end{array}
       \right)=\left(
         \begin{array}{c}\vspace{.2cm}
           y_1 \\\vspace{.2cm}
           y_2 \\\vspace{.2cm}
           \vdots \\
           y_k \\
         \end{array}
       \right) \mod (q-1).
\ee

Denote the coefficient matrix of above linear system as $A$. Then by a result in \cite{dickson}
there exist $L\in SL(k,\mathbb{Z}), R\in SL(n,\mathbb{Z})$, $D=\textmd{diag}\{s_1,s_2,\ldots,s_r\}$ and $s_i\mid s_{i+1}$ for $1\leq i\leq r-1$, such that
\be\label{smith}
LAR=\left(
        \begin{array}{ll}\vspace{.2cm}
          D                 ~~& 0_{r\times (n-r)} \\
          0_{(k-r)\times r} ~~& 0_{(k-r)\times (n-r)} \\
        \end{array}
      \right)=:J.
\ee
Denote $\x=(x_1,\ldots,x_n)^T,\y=(y_1,\ldots,y_k)^T$ and $\tilde{\x}=R^{-1}\x=(\tilde{x}_1,\ldots,\tilde{x}_n)^T, \tilde{\y}=L\y=(\tilde{y}_1,\ldots,\tilde{y}_k)^T$, then (\ref{eq10}) is equivalent to
\be\label{eq11}
J\tilde{\x}=\tilde{\y}\mod (q-1).
\ee
Equation (\ref{eq11}) has a solution if and only if
\be
\gcd(s_i,q-1)\mid \tilde{y}_i,~\textmd{for}~i=1,2,\ldots,r;~\textmd{and}~\tilde{y}_j=0~\textmd{for}~j=r+1,\ldots,k.
\ee

In (\ref{smith}), $J$ is called the \emph{Smith normal from} of $A$. As we can see $s_1,\ldots,s_r$ are totaly determined by $A$. They form the \emph{invariant factors} of $A$. If $\gcd(s_r,q-1)=1$ and $r=k$, then any solution of (\ref{eq9}) will give a solution of (\ref{eq8}). Combining the above analysis, we have

\bp
Let $\mathbb{F}_q$ be a finite field, $A=(a_{ij})$ be a matrix of size $k\times n$ and its rank equals $k$. Suppose $s_1,s_2,\ldots,s_k$ are the invariant factors of $A$ and $\gcd(s_k,q-1)=1$.
Let $d_i$ be the greatest common divisor of elements in the $i$-th row of $A$ for $1\leq i\leq k$.

(1). Any solution $\y$ of (\ref{eq9}) will recover a solution $\x$ of equation (\ref{eq8}). And the costs from $\y$ to $\x$ is $O(\emph{poly}(k,n))$.

(2). If there is a $1\leq i\leq k$, such that $d_i=1$, then there is a quantum algorithm find the solution of polynomial equation (\ref{eq8}) in time $O(\emph{poly}(k,n,\log q))$.
\ep

So the study of equation (\ref{eq7}) has its own importance. It might be a starting point to study the general polynomial equation in finite field.

\section{Conclusion}

In this article, we have considered three different types of equations. Each equation has its own character, and provides more interesting problems which deserve to study in the future.


\begin{thebibliography}{9}

\small

\bibitem{aaronson} Aaronson, S., \emph{The limits of quantum computers}, Science American, 62-69, 2008.

\bibitem{adleman} Adleman, L. and Manders, K., \emph{Reducibility, randomness, and intractability (Abstract)}, Proc. 9th Ann. ACM Symp. on Theory of Computing, Association for Computing Machinery, New York, 151-163. 1977.

\bibitem{berlekamp} Berlekamp E.R., McEliece R.J. and van Tilborg H.C., \emph{On the inherent intractability of certain coding problems}. IEEE Trans Inf Theory 24(3): 384-386, 1978.

\bibitem{brassard} Brassard, G., H{\o}yer, P. and Alain Tapp, A., \emph{Quantum cryptanalysis of hash and claw-free functions}, ACM SIGACT News 28(2): 14-19, 1997.

\bibitem{chen} Chen, J.X., Childs, A.M. and Hung, S.H., \emph{Quantum algorithm for multivariate polynomial interpolation}, arXiv:1701.03990.

\bibitem{childs} Childs, A.M. and van Dam, W., \emph{Quantum algorithms for algebraic problems}, Reviews of Modern Physics 82, 1-52, 2010.

\bibitem{childs2} Childs, A.M., van Dam, Hung, S.H. and Shparlinski, I.E., \emph{Optimal quantum algorithm for polynomial interpolation}, Proceedings of the 43rd International Colloquium on Automata, Languages, and Programming (ICALP 2016), pp. 16:1-16:13, 2016.

\bibitem{dickson} Dickson L.E., \emph{History of the Theory of Numbers}, Volume 2, G.E. Stechert \& Co., New York, 1934.

\bibitem{dubrois} Dubrois, J. and Dumas, J.G., \emph{Efficient polynomial time algorithms computing industrial-strength primitive roots}, Information Processing Letters 97(2): 41-45, 2006.

\bibitem{garey} Garey, M.R. and Johnson, D.S., \emph{Computers and Intractability - A Guide to the Theory of NP-Completeness},  W. H. Freeman; 1st Edition edition, 1979.

\bibitem{grover} Grover L.K., \emph{A fast quantum mechanical algorithm for database search}, Proceedings, 28th Annual ACM Symposium on the Theory of Computing, pp. 212-219, 1996.

\bibitem{harrow} Harrow, A.W., Hassidim, A. and Lloyd, S., \emph{Quantum algorithm for solving linear systems of equations}, Phys. Rev. Lett. 15(103), pp. 150502, 2009.

\bibitem{hua} Hua, L-K., \emph{Introduction to Number Theory}, Springer, 1982.

\bibitem{manders} Manders, K. and Adleman, L., \emph{NP-complete decision problems for binary quadratics}, Comput. System Sci. 16, 168-184, 1978.

\bibitem{mullen} Mullen, G.L. and Panario, D., \emph{Handbook of finite field}, Chapman and Hall/CRC; 1 edition, 2013.

\bibitem{shor} Shor, P.W., \emph{Polynomial-Time Algorithms for Prime Factorization and Discrete Logarithms on a Quantum Computer}, SIAM J. Comput., 26(5): 1484-1509, 1997.

\bibitem{dam} van Dam, W. and Shparlinski, I., \emph{Classical and quantum algorithms for exponential congruences}. Proceedings of TQC 2008, pg. 1-10, 2008.

\end{thebibliography}
\end{document}